# Advanced modelling of a moderate-resolution holographic spectrograph

Eduard Muslimov,[1,2*] Gennady Valyavin,[3] Sergei Fabrika,[3,4] Nadezhda Pavlycheva,[2]

[1] Aix Marseille Univ, CNRS, LAM, Laboratoire d'Astrophysique de Marseille, Marseille, France, 38 rue Joliot-Curie, Marseille 13388, France
[2] Kazan National Research Technical University named after A.N. Tupolev –KAI, 10 K. Marx, Kazan 420111, Russian Federation
[3] Special Astrophysical Observatory, Russian Academy of Sciences, Nizhnii Arkhyz 369167, Russian Federation
[4] Kazan Federal University, 18 Kremlevskaya, Kazan 420008, Russian Federation
*Corresponding author: eduard.muslimov@lam.fr

In the present article we consider an accurate modeling of spectrograph with cascade of volume-phase holographic gratings. The proposed optical scheme allows to detect spectra in an extended wavelength range without gaps providing relatively high spectral resolution and high throughput. However, modeling and minimization of possible cross-talks between gratings and stray light in such a scheme represents a separate task. We use analytical equations of coupled wave theory together with rigorous coupled wave analysis to optimize the gratings parameters and further apply the latter together with non-sequential raytracing algorithm to model propagation of beams through the spectrograph. The results show relatively high throughput up to 53% and absence of any significant cross-talks or ghost images even for ordinary holograms recorded on dichromated gelatin. © 2016 Optical Society of America

# 1. INTRODUCTION

It is commonly known that thick volume-phase gratings are characterized by high diffractive efficiency in the maximum together with high spectral selectivity [1]. These properties were used a few times for building of spectral instruments with an enhanced performance [2-5].

We proposed to use a cascade of volume-phase holographic (VPH) grating in a moderate-resolution spectrograph with improved throughput. In such a spectrograph a number of gratings are mounted in a parallel beam one after another. Each of them has very high diffraction efficiency (DE) in a specified wavelength range. Outside of it the DE is negligible, so almost all the radiation passes to the $0^{th}$ order and the grating operates as a plane-parallel plate. With use of this principle it is possible to obtain a spectral image consisting of a few lines. The lines can have equal linear dispersion and be centered due to a proper choice of grooves frequency and inclination in the tangential plane for each of the grating. A necessary spacing between the lines is provided by inclination of the gratings in the sagittal plane.

On the previous stages of the researches we developed two versions of such spectrograph for the visible domain 430-680 nm [6,7]. It was demonstrated that the spectral resolution of the full-size scheme is 0.082-0.124 nm, i.e. the spectral resolving power is 5243-7906. The corresponding values for the reduced version of spectrograph are 0.125-0.330 nm and 1553-5124, respectively. However, for the total throughput and its spectral dependence only approximate estimations were performed.

At the same time, throughput and possible presence of non-working diffraction orders in the image is the key question for the proposed design, which can limit its implementation and applications. So the main goal of the present research is to model accurately diffraction efficiency and radiation splitting between diffraction orders for the gratings in real scheme and quantitate of entire spectrograph.

The paper is organized as follows. In section 2 we provide a short overview of the optical scheme under consideration and specify the initial data for modelling. Section 3 includes description of the modelling strategy and indicates the used methods. Further, section 4 describes DE modeling. Section 5 presents results of raytracing through entire spectrograph scheme with use of data obtained on the previous stage. In section 6 we discuss the obtained results and possible use of the developed spectrograph with a special accent on astronomical applications.

# 2. OPTICAL DESIGN OVERVIEW

We consider the reduced and simplified version of the spectrograph scheme only [7]. The first reason for this decision is that this scheme is intended for creation of a lab prototype. So it is important to have accurate modelling results to predict the experimental performance. Further, the scheme consists of simpler optical elements, so it is easier to model them. Namely, in contrast with the full-scale spectrograph, there are n wedges in this version.

The reduced optical scheme of spectrograph operates in the visible domain 430-680 nm. It is divided to three sub-ranges: 430-513, 513- 597 and 597-680 nm. The dispersive unit consists of three VPH gratings working in the corresponding sub-ranges. Each of the gratings is imposed on a plane-parallel plate made of BK7 glass and protected by a cover glass (the substrate thickness is 2.6 mm and the cover glass thickness is 2.2 mm). The length of each spectral image is 20 mm (so the reciprocal linear dispersion is 4.15 nm/mm). The limits on lines spacing in the spectral are 1.5-4 mm. The gratings are mounted in collimated beams, so two identical commercial Tessar-type lenses are used as the collimator and camera. Each of the lenses has focal length of 135 mm and F/# of 2.8. The actual entrance aperture is assumed to be decreased to F/# = 4.

It was shown that the scheme provides spectral resolution of 0.125-0.203 nm; 0.125-0.330 nm and 0.125-0.151 nm in three subranges, specified above, respectively. So the spectral resolving power is 1553-5124.

The optical scheme general view is presented on Fig.1. Axes notation necessary for further explanations is provided on this figure as well.

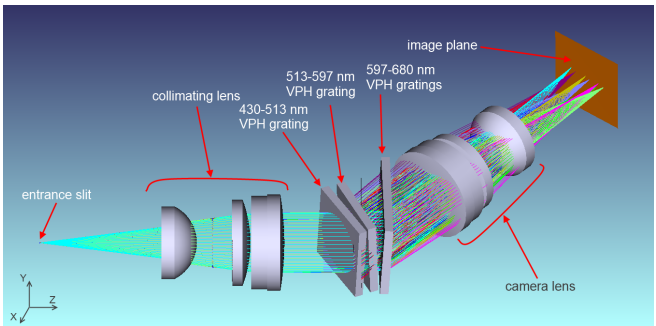

Fig. 1. General view of the spectrograph optical scheme.

The initial data for DE modelling consists of spatial frequency, working spectral range and angular position in respect to the incident beam for each grating. This data is summarized in Table 1. Note, that the angular positions are given via direction cosines on the holographic layer, i.e. they are measured after the substrate.

Table 1. Initial data on the spectrograph VPH gratings

| Grating # | 1 | 2 | 3 |
|---|---|---|---|
| Working range (nm) | 430-513 | 513-597 | 597-680 |
| Grooves frequency (mm$^{-1}$) | 1726 | 1523 | 1205 |
| Direction cosine with X axis | -0.0341 | 0.0196 | -0.0413 |
| Direction cosine with Y axis | -0.3116 | -0.2562 | -0.0202 |
| Direction cosine with Z axis | 0.9496 | 0.9664 | 0.9989 |

Extraction of these data from the spectrograph scheme is the first step of our modelling algorithm, which is described further.

## 3. MODELLING STRATEGY

The modeling algorithm is presented in a form of block diagram on Fig.2 including the main modeling steps and parameters defined on each of them. The methods and software used as well as other details of each stage are described below.

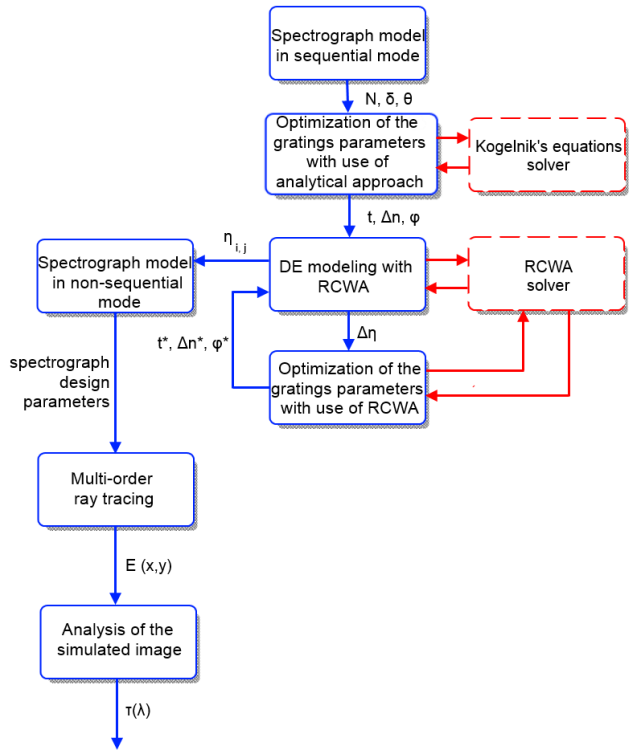

Fig. 2. Block diagram of the spectrograph modelling algorithm.

After receiving of the gratings parameters they should be converted to a form suitable for further calculations. Namely, spherical angles ($δ, θ$) are used to describe the incident beam instead of the direction cosines.

On the next step we use analytical expressions of the Kogelnik's coupled wave theory [8] to optimize the holographic layers' parameters. In each case the fringes inclination angle $φ$ is defined directly by the Bragg condition. The thickness $t$ and refraction index modulation $Δn$ are found by local minimization of a simple merit function in area of technologically feasible parameters values.

These values are transferred to software, which models diffraction on a volume grating with use of numerical rigorous coupled wave (RCWA) algorithm. We use *reticolo* software, developed by J.-P. Hugonin and P. Lalanne [9, 10]. It is very important to emphasize here, that in our case a check with accurate numerical method is extremely necessary. Firstly, the gratings in optical scheme are turned around the Y axis to divide the image lines in the vertical direction. It causes conical diffraction, which is difficult to model with Kogelnik's theory.  Secondly, in the cascaded mounting, which we use, energy splitting between diffraction orders becomes a critical factor. The analytical coupled wave theory uses simplifying assumptions relating to the diffraction orders, so it can give an impropriate results.

We should emphasize that the first three steps were considered in our previous studies [6, 7] and here they serve as a starting point for further optimization and modelling.

Accounting for the mentioned difference between the analytical and numerical grating models (and corresponding DE difference $\Delta \eta$) on the next step we optimize the grating parameters with use of RCWA. The values obtained before are used as a starting point. Here we have to set fringes inclination angle as a free optimization parameter in order to consider for the conical diffraction.

Then we repeat numerical modelling for the corrected grating parameters $t^*$, $\Delta n^*$ and $\varphi^*$ and obtain real DE values for all the gratings and diffraction gratings at different wavelengths. To limit the amount of data and modelling time the data array transferred to raytracing software contains DE values for 11 wavelengths covering the working range of 430-680 nm. The orders of diffraction under consideration are $+1^{st}$, $0^{th}$ and $-1^{st}$. All the computations are performed for unpolarized light.

On the raytracing stage we account for all the optical components of the spectrograph. Standard non-sequential mode of Zemax software is used. Rays are split on refractive and diffractive surfaces. A simple macro performs the analysis for a number of wavelengths with data acquisition on the detector in a loop.

Finally, the simulated spectral image is analyzed separately. Each monochromatic slit image is extracted and irradiance is integrated over it is square to compute the flux. Thus the total throughput of the optical scheme is defined.

## 4. DIFFRACTION EFFICIENCY MODELLING AND OPTIMIZATION

At first, the gratings DE is optimized and modelled with use of Kogelnik's theory. The target DE spectral dependence is a rectangular function

$$\eta_{tar}(\lambda) = \begin{cases} 1, \lambda_1 < \lambda < \lambda_2; \\ 0, otherwise. \end{cases} \quad (1)$$

where ($\lambda_1$; $\lambda_2$) is the corresponding spectral subrange.

So the merit function represents simply a RMS deviation of a real DE from this target

$$f_m(t, \Delta n) = \int_{400}^{800} (\eta(t, \Delta n, \lambda) - \eta_{tar}(\lambda))^2 d\lambda \quad (2)$$

This merit function is optimized locally in a region of technologically realizable parameters [11]. We suppose that the holographic gratings are recorded on dichromated gelatin (DCG) layers with use of common technology, which implies a sinusoidal or quasi-sinusoidal variation of the refraction index. Thus, the refraction index modulation changes between 0.0005 and 0.05 and the holographic layer thickness can possess the values from 5 to 35 µm [12, 13]. The dispersive properties of DCG and glass substrate are taken into account [14, 15].

Example of result obtained during the optimization is shown on Fig.3, which represents analytical DE distribution for the first grating.

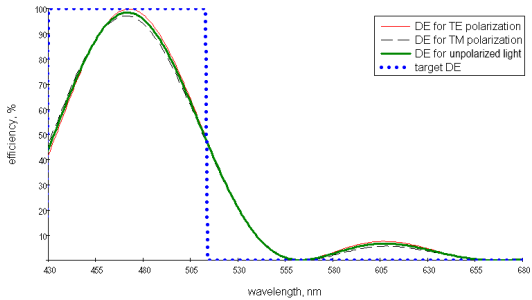

Fig. 3. Spectral dependence of DE for the first grating obtained with analytical method.

A few notes should be made here. Firstly, the grating has significant residual efficiency outside of the working subrange, including the secondary maximum. This is an unavoidable shortcoming of ordinary sine-profiled gratings and the main reason to perform the accurate modeling of the scheme. Secondly, polarization dependence is relatively weak, so the DE values for TE and TM polarization states differs for only 3%.  The corresponding values for the second ('green') and third ('red') gratings are 6.3 and 8.6 % respectively. Hereafter we provide data for unpolarized light only.

The gratings parameters found with the analytical method are listed in Table 2.

Table 2. Gratings parameters optimized with use of analytical equations of coupled wave theory

| Grating # | 1 | 2 | 3 |
|---|---|---|---|
| Holographic layer thickness (µm) | 10 | 12.8 | 19.1 |
| Refraction index modulation | 0.023 | 0.021 | 0.014 |
| Fringes inclination angle | 3.305° | 6.143° | 12.602° |
| First recording angle | 33.025° | 34.485° | 40.776° |
| Second recording angle | -21.914° | -14.125° | 0.690° |

For the gratings with optimized parameters the DE distribution was computed by the RCWA code. The results are presented on Fig.4. It is clear that in comparison with predictions of the analytical theory actual DE curves are lower and blueshifted. Moreover the blueshift is proportional to the grating tilt around the Y axis.

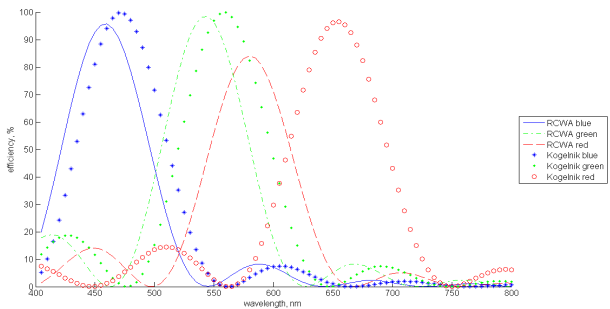

Fig. 4. Comparison of DE curves obtained with the analytical and numerical methods.

Gratings with such DE distribution are inappropriate for use in the spectrograph. So the optimization was repeated. The merit function (2) remains almost the same, but the DE is calculated with use of RCWA method and the fringes inclination angle is set to be a free optimization parameter. Results of DE spectral dependence computation for the corrected parameters are shown on Fig. 5.

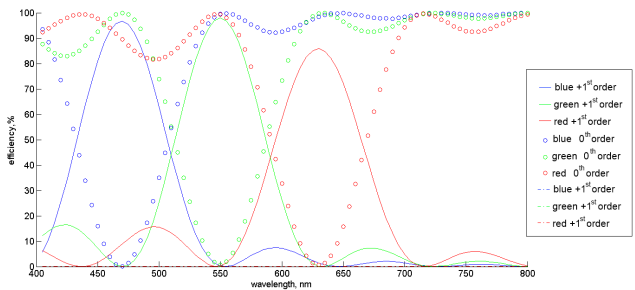

Fig. 5. Results of accurate numerical modelling of the VPH gratings with corrected parameters.

In this computation -1$^{st}$, 0$^{th}$ and +1$^{st}$ diffraction orders are taken into account. One can see that contribution of the -1$^{st}$ order is negligible.

The corrected values of the gratings parameters are listed in Table 3.

Table 3. Gratings parameters optimized with use of RCWA

| Grating # | 1 | 2 | 3 |
|---|---|---|---|
| Holographic layer thickness (µm) | 10.4 | 13.2 | 18.1 |
| Refraction index modulation | 0.021 | 0.019 | 0.018 |
| Fringes inclination angle | 3.701° | 6.358° | 13.814° |
| First recording angle | 33.731 ° | 34.876 ° | 43.191 ° |
| Second recording angle | -21.281° | -13.794° | 2.486 ° |

Further the obtained data about DE values and orders splitting is recorded to a file for use in raytracing.

## 5. EVALUATION OF ENTIRE SPECTROGRAPH THROUGHPUT

Using the data about energy splitting between orders and DE spectral dependence a raytracing procedure is performed. The spectrograph optical system is modeled in Zemax non-sequential mode. It is supposed that all the refracting surfaces have anti-reflecting coatings with residual reflectivity ≤1% and the side cylindrical surface of each lens is covered by a black absorbing paint. The raytracing is performed in a loop for 11 fixed wavelengths.

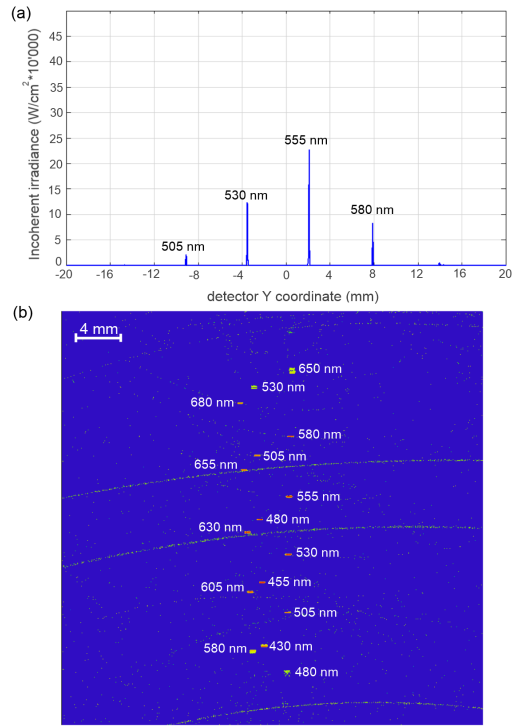

Fig. 6. Results of raytracing in non-sequential mode: (a) irradiance across 513-597 spectral line in linear scale; (b) full-frame simulated image in log scale.

Fig. 6 represents the raytracing results. Note, that log scale is used for the full-frame image to emphasize residual cross-talks and stray light. In addition, for simplicity reasons it was supposed that each monochromatic incident beam has power of 100 W. One can see, that due to intersection of the DE curves weak spectral images corresponding to wavelengths outside of the calculated range appear in each of the lines. They decrease total throughput, but cannot embarrass the spectrum detection and lines identification. These excessive spectral images concentrate 60.5% of flux which was not directed to the registered spectra or 4.5% of total flux and can be completely eliminated by means of calibration. The residual stray light is introduced mostly by refraction on the lenses edges; in comparison with the signal it is smaller for a few orders of magnitude, so it also can not affect the measurement process.

The obtained simulated image was processed. Monochromatic image of the entrance slit at each of the 11 wavelengths was extracted and integrated over the area to calculate total flux. The separated images are shown on Fig.7. Besides, the spectrograph aberrations can be seen on these plots.

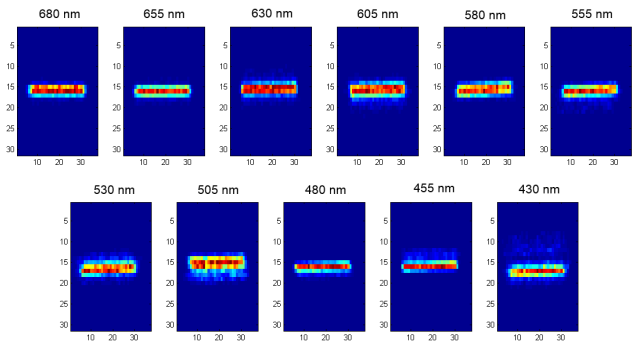

Fig. 7. Simulated monochromatic images of the entrance slit.

Finally, the flux values were divided by the incident flux and the total throughput was defined as a function of wavelength. To provide a scale to compare with, we repeated the same calculations for an ideal case, when the DE curves are described by rectangular functions (1). The computation results are presented on Fig.8. Throughput for real gratings reaches 53% with drops up to 12% and 22.5% at the borders of sub-ranges. Irregularity of the transmittance graph for ideal case indicates on presence of other throughput change sources, like vignetting and variation of the incidence angle across a lens.

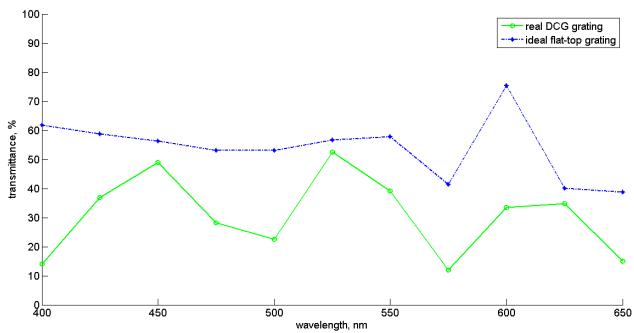

Fig. 8. Total spectrograph throughput spectral dependence.

Therefore, the total throughput values computed with the advanced modeling algorithm are generally lower than that obtained before using an approximate estimation [7]. However, these values are substantially higher than throughput of existing moderate and high resolution spectrographs.

It is necessary to state here that we considered only the simplest case of a VPH grating with sine profile of the refraction index modulation. Use of modern holographic materials and techniques as well as improvement of the recording layout can allow to manage the DE shape for individual grating. With this condition the total throughput would approach the limit shown for ideal case. We keep this topic for future research. Nevertheless, the obtained results show relatively high throughput exceeding those for other classes of spectrographs.

## 6. CONCLUSIONS

We considered an optical scheme of spectrograph with cascade of volume-phase holographic gratings. The spectrograph operation principle uses spectral selectivity of VPH gratings to create a spectral image consisting of 3 lines with moderately high dispersion and spectral resolution. Its' small-sized version provides spectral resolving power of 1553-5124 in the visible domain 430-680 nm.

It was shown that the gratings inclination in sagittal plane, which is necessary to separate the image lines, causes conical diffraction and subsequently a considerable difference of real diffraction efficiency from its values calculated with an analytical approach. So the holographic layer parameters for each grating were optimized with use of numerical RCWA method.

Finally, using the RCWA modelling results obtained after the optimization, we performed raytracing through non-sequential spectrograph model. It was shown that the scheme is free of ghost images and the stray light is negligible, while some residual cross-talks present. As a result, the spectrograph total throughput reaches 53% in the peak and has noticeable drops on the edges of sub-ranges.

The results show that the developed scheme has a number of advantages in comparison with existing spectrographs between low-resolution instruments (resolution less than 1000 for the whole visible range, a total transmission higher than

50%) and high-resolution echelle spectrographs (resolution higher than 10000, a total transmission less than about 10%). Our cascade-VPH scheme has advantages in both above mentioned types of spectrographs, it can achieve a resolution at least 5000 and a total transmission up to ~60%. This solution provides comparatively high resolving power and conserves a typical for echelle-spectroscopy wide spectral region with much higher efficiency.

It is well-known fact, that efficiencies of even the best existing echelle-spectrometers HARPS [16] and PEPSI [17] still not exceed 10 percent while standard moderate and low-resolution spectrographs are 5-7 times more efficient (for example, FOCAS [18]). This makes the use of the cascade-VPH spectrograph more effective in studies of faint, and/or low-contrast objects. When the best moderate-resolution spectrographs have spectral resolution of 1000 (at higher resolution these can not cover all the visible range) that means equivalent spectral resolution of 300 km/s. In such case is it impossible to distinguish between nebulae, stellar winds, or accretion disks surrounding black holes, because the nebulae have a dispersion velocity less than 50 km/sec, stellar winds or accretion disks have the dispersion higher than 100 - 200 km/sec. That is crucially new to have spectral resolution higher than 5000 (60 km/sec), and therefore our solution is able to considerably contribute to modern areas of astrophysics. Among these are studies of evolved massive stars in the Galaxy and nearby galaxies [19], X-ray sources in the Galaxy and ultra-luminous X-ray sources in other galaxies [20]; young stellar clusters in the local universe [21]; a search for signatures of reflected light from "hot-Jupiter" exoplanets [22]; spectral studies of variable magnetic white dwarf stars [23-25] and others.

It is also evident that an efficient moderate-resolution spectrograph operating in wide spectral range is required not only for the above mentioned fields, but also for many currently important fields of astrophysics. We would like to notice that a number of astrophysical studies for the spectrograph can be increased with broadened of wavelength region, especially toward shorter wavelengths. We will consider such a possibility in a special further study.

On the next stage of the research the spectrograph key performance parameters should be verified experimentally. It implies building and alignment of the spectrograph prototype and measurement of its' spectral resolution and total throughput, as well as estimation of stray light with different sources. Altogether, it represents an extensive experimental program, which will be considered separately.

**Funding Information: Russian Science Foundation** (14-50-00043).

**Acknowledgments**. We thank J.-P. Hugonin and Ph. Lalanne for provision of the *reticolo* software used in this work.